\documentclass{aa}
\usepackage{graphicx}
\usepackage{txfonts}

\usepackage[]{amsmath}     
\usepackage{natbib}        
\bibpunct{(}{)}{;}{a}{}{,} 
\usepackage{color}

\begin{document}

\title{Separating gas-giant and ice-giant planets \\ 
  by halting pebble accretion}
\author{
  M. Lambrechts \inst{1} \and 
  A. Johansen\inst{1} \and
  A. Morbidelli \inst{2}
} 

\institute{Lund Observatory, Department of Astronomy and Theoretical
Physics, Lund University, Box 43, 22100 Lund, Sweden\\
\email{michiel@astro.lu.se}
\and
Dep. Lagrange, UNSA, CNRS, OCA, Nice, France
}
   
\date{Received --- ; accepted ---}

\abstract{
In the Solar System giant planets come in two flavours: `gas giants' (Jupiter
and Saturn) with massive gas envelopes and `ice giants' (Uranus and Neptune)
with much thinner envelopes around their cores.
It is poorly understood how these two classes of planets formed.
High solid accretion rates, necessary to form the cores of giant planets
within the life-time of protoplanetary discs, heat the envelope and prevent
rapid gas contraction onto the core, unless accretion is halted.
We find that, in fact, accretion of pebbles (\mbox{$\sim$\,cm-sized}
particles) is self-limiting: when a core becomes massive enough it carves a
gap in the pebble disc. 
This halt in pebble accretion subsequently triggers the rapid collapse of
the super-critical gas envelope.
As opposed to gas giants, ice giants do not reach this threshold mass and can
only bind low-mass envelopes that are highly enriched by water vapour
from sublimated icy pebbles.
This offers an explanation for the compositional difference between gas
giants and ice giants in the Solar System.
Furthermore, as opposed to planetesimal-driven accretion scenarios, our model
allows core formation and envelope attraction within disc life-times, provided
that solids in protoplanetary discs are predominantly in pebbles.
Our results imply that the outer regions of planetary systems, where the mass
required to halt pebble accretion is large, are dominated by ice giants and
that gas-giant exoplanets in wide orbits are enriched by more than 50 Earth masses of solids.
}


\maketitle

%
%

\section{Introduction}

In the core accretion scenario \citep{Pollack_1996}, giant planets form by
attracting a gaseous envelope onto a core of rock and ice.
This theory is supported by the large amount of heavy elements -- elements
with atomic number above He -- found in the giant planets in our Solar System
\citep{Guillot_2005}.
Further evidence is provided by the observed dependence of giant exoplanet
occurrence on the host star metallicity, which is a proxy for the dust mass
enrichment of the protoplanetary disc \citep{Fischer_2005,Buchhave_2012}.

However, from a theoretical perspective it is poorly understood how the core accretion scenario
could have taken place, if the cores grew by accretion of km-sized
planetesimals and their fragments. 
Protoplanetary disc life-times range from $\sim$$3$\,Myr
\citep[][]{Haisch_2001,Soderblom_2013} to possibly as long as $\sim$$6$\,Myr
\citep{Bell_2013}. 
This is much shorter than the time needed to grow cores to completion in
numerical simulations \citep{Levison_2010} of discs with solid surface
densities comparable to the minimum mass solar nebula
\citep[MMSN,][]{Hayashi_1981}.
Additionally, the gaseous envelope grows only slowly on Myr time-scales, because of the continued heating by accretion of remnant planetesimals, even
after clearing most of its feeding zone \citep{Pollack_1996, Ikoma_2000}.
Therefore, planets with gaseous envelopes are difficult to form by planetesimal
growth within $\sim$$10$\,Myr, especially outside the current orbit of Jupiter
(5\,AU), where core growth timescales rapidly increase \citep{Dodson_2009}.

As a result, protoplanetary discs with strongly enhanced solid surface
densities in planetesimals \citep[exceeding the MMSN by a factor
10,][]{Kobayashi_2011} have been proposed in order to form the cores of the
giant planets. 
For the gas giants, planetesimal accretion is then halted artificially, or the
opacity in the envelope is lowered, in  order to reduce the envelope attraction
timescale \citep{Hubickyj_2005}.
The ice giants are envisioned to remain small, because the protoplanetary gas
disc dissipates during slow envelope growth \citep{Pollack_1996,Dodson_2010}.

In this paper, we investigate the attraction of the gaseous envelope
when growth occurs by the accretion of \emph{pebbles}, as opposed to
planetesimals.
Pebble accretion rates are sufficiently high to form the cores of giant planets
in less than $1$ million years, even in wide orbits \citep{Lambrechts_2012}.
Previous studies \citep{Johansen_2010, Ormel_2010,
Bromley_2011, Lambrechts_2012,Morbidelli_2012} demonstrate that this is the result of gas drag operating on pebbles, which dramatically
increases the accretion cross section (Section\,2).
The rapid accretion of pebbles leads to high accretion luminosities that
support a growing gaseous envelope around the core (Section\,3). 
We proceed by calculating the critical core mass, the lowest mass for which a
core can no longer sustain the hydrostatic balance of the proto-envelope. 
The critical core masses we find are on the order of $\approx$$100$ Earth
masses (M$_{\rm E}$), too large compared to the inferred core masses of the
gas giants in the Solar System.
Fortunately, we find that there is a threshold mass already around $20$\,M$_{\rm
E}$, where the core perturbs the gas disc and halts the accretion of pebbles,
which initiates the collapse of the envelope before the critical core mass is
reached (Section\,4).
This threshold mass is reached by the cores of the gas giants, but not by the
ice giants in wider orbits. 
By combining our calculations of the pebble isolation mass and the critical
core mass as function of the envelope enrichment, we can make estimates of the
bulk heavy element content of the giant planets. 
We find a good agreement with the composition of the giant planets in the Solar
System (Section\,5). 
We also discuss the implications of our model on the occurrence and composition
of giant exoplanets (Section\,6). 
Finally, we briefly summarize our work (Section\,7).

\section{Pebble accretion}

\begin{figure}
  \centering
  \includegraphics{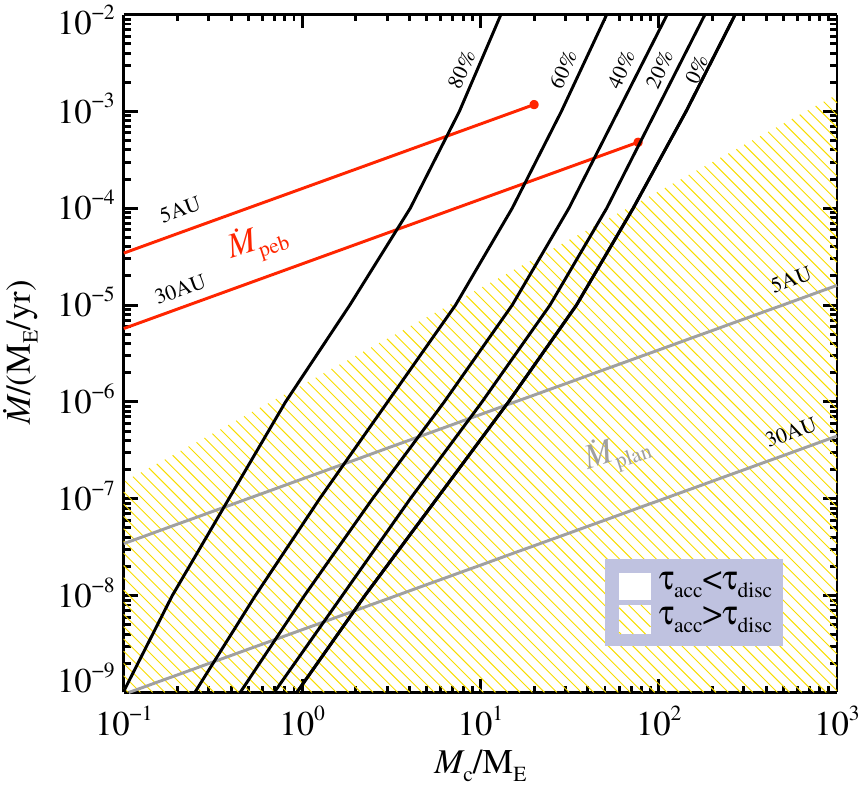}
  \caption{
  Pebble accretion rates (red), planetesimal accretion rates (grey) and minimal
  accretion rates required to sustain a stable gas envelope (black), as
  function of the core mass.
  The latter `critical' curves are nearly independent of orbital radius between
  $5$-$30$\,AU, but depend strongly on the opacity
  (Appendix\,\ref{sec:composition}) and on the level of envelope pollution by
  sublimation of icy pebbles. 
  Labels at the top of the figure indicate H$_2$O pollution of the atmosphere
  as a percentage with respect to pure H/He nebular gas, corresponding to the
  fraction $1-\beta$ from Eq.\,(\ref{eq:lum_acc}). 
  The critical core mass to collapse the gas envelope can be found at the
  intersection of an accretion curve with a critical curve.  
  Accretion rate curves that fall in the yellow dashed region are too slow  to
  form the cores of the giant planets before the dissipation of the gas disc
  ($\tau_{\rm acc} = M_{\rm c}/\dot M_{\rm c} > \tau_{\rm disc} = 2$\,Myr).
  Red circles mark the mass above which pebble accretion is halted
  (Eq.\,\ref{eq:isomass}) and the gravitational collapse of the gas envelope is
  triggered.
}
  \label{fig:mvsmdotcrit}
\end{figure}

The pebble accretion scenario, as outlined in \citet{Lambrechts_2012}, starts
with the growth of pebbles from the initial grains embedded in the
protoplanetary disc (with sizes $\approx$\,$\mu$m) by collisions
\citep{Birnstiel_2012} or through sublimation and condensation cycles around ice
lines \citep{Ros_2013}.  
A fraction of the population of pebbles that drift towards the host star form
dense swarms that subsequently collapse under self-gravity to create
planetesimals $100$-$1000$\,km in size. 
Such concentrations can occur through the streaming instability, driven by the
mutual drag between particles and gas \citep{Youdin_2005, Youdin_2007,
Johansen_2007}, or for example through the presence of vortices
\citep{Barge_1995} or pressure bumps \citep{Whipple_1972}. 
A further discussion can be found in the reviews by \citet{Chiang_2010} and
\citet{Johansen_2014}.
Finally, the largest planetesimals can act as the seeds of the planetary cores
which grow by rapidly sweeping up the remaining pebbles 
\citep{Lambrechts_2012}.

We consider here cores that grow dominantly by the accretion of particles with
radii of approximately mm-cm.
Particle sizes can be expressed as function of the gas drag time-scale ($t_{\rm
f}$) and Keplerian frequency $\Omega_{\rm K}$ in terms of the Stokes number
\begin{align}
  \tau_{\rm f} =\Omega_{\rm K}t_{\rm f} =\frac{\rho_\bullet R}{\rho H},
  \label{eq:stokes}
\end{align}
where $\rho_\bullet$ is the solid density, $R$ the particle radius, $\rho$ the
midplane gas density and $H$ the local gas scale height of the disc. 
Small dust particles ($\tau_{\rm f} \ll 1$) are thus strongly coupled and
comoving with the gas, while much larger objects ($\tau_{\rm f} \gg 1$) are
only weakly affected by gas gas drag.
In the outer parts of the MMSN, in the region with semi-major axis $a$ between
$5$ to $30$\,AU, particle sizes between mm and cm correspond to $\tau_{\rm f}
\approx 0.01-0.1$.

The seeds of the planetary cores 
accrete from the full scale height of pebbles at a rate
\begin{align}
  \dot M_{\rm c} = 2 
  r_{ \rm H} \Sigma_{\rm p} v_{\rm H},
  \label{eq:Mdot_small_pebbles}
\end{align}
\citep{Ormel_2010, Lambrechts_2012}.
Here, $\Sigma_{\rm p}$ denotes for the surface density in pebbles and $v_{\rm
H}=r_{\rm H}\Omega_{\rm K}$ is the Hill velocity at the Hill radius $r_{\rm H} =
\left[ GM_{\rm c}/(3\Omega_{\rm K}^2) \right]^{1/3}$, with $G$ the gravitational
constant.
Particles entering the Hill sphere have a crossing time, $\tau_{\rm c}\sim
r_{\rm H}/v_{\rm H}$, comparable to the orbital time-scale. 
Gas drag operates on pebbles on similar time-scales, leading to their accretion
by the core.
This accretion rate does not depend on the particle size between $\tau_{\rm f}
= 0.1-10$, but moderately decreases $\propto (\tau_{\rm f}/0.1)^{2/3}$ for
particles below $\tau_{\rm f}=0.1$ \citep{Lambrechts_2012}.
We have assumed in Eq.\,\ref{eq:Mdot_small_pebbles} that the particle
scale height is smaller than the Hill radius of the core, which is valid when
core masses are larger than
\begin{align}
  M_{\rm c, 2D} \approx 0.19 \left( \frac{\tau_{\rm f}}{0.1} \right)^{-3/2} 
  \left( \frac{\alpha_{\rm t}}{10^{-3}} \right)^{3/2} 
  \left( \frac{a}{10\,{\rm AU}} \right)^{-3/4}\,{\rm M}_{\rm E}.
  \label{eq:Mc_2D}
\end{align}
Here we have taken for simplicity an MMSN model with a particle scale height
given by $H_{\rm p}/H = \sqrt{\alpha_{\rm t}/\tau}$ \citep{Youdin_2007b}, where
$\alpha_{\rm t}$ is the turbulent diffusion parameter. 
Low particle scale heights are expected in dead-zones and discs where angular
momentum transport occurs primarily through disc winds \citep{Turner_2014}.
From particle stirring alone, scale heights of $H_{\rm p}/H\approx 0.01$ are
expected \citep{Bai_2010}. 
In the MMSN, the accretion rate when $H_{\rm p} < r_{\rm H}$ translates into
\begin{align}
  \dot M_{\rm c} &= 80
  \left( \frac{M_{\rm c}}{1M_\oplus} \right)^{2/3} 
  \left(\frac{a}{10{\rm\,AU}}\right)^{-1} 
  \frac{{\rm M}_{\rm E}}{\rm 10^6\,yr},
  \label{eq:pebble_acc_rate} 
\end{align}
which is illustrated in Fig.\,\ref{fig:mvsmdotcrit} (red curves).

The growth of the core is driven by the radial drift of pebbles through the
protoplanetary disc.
Because of gas drag robbing pebbles of angular momentum, they spiral towards the star with a velocity
\begin{align}
  v_{\rm r} \approx - 2 \tau_{\rm f} \eta v_{\rm K},
  \label{eq:rad_vel}
\end{align}
for particles with $\tau_{\rm f} < 1$ \citep{Weidenschilling_1977, Nakagawa_1986}. Here $v_{\rm K}$ is
the Keplerian velocity and 
\begin{align}
  \eta &= -\frac{1}{2} \left( \frac{H}{a} \right)^2
           \frac{\partial \ln P}{\partial \ln a} 
       \approx 0.0015 \left( \frac{a}{\rm AU} \right)^{1/2}
  \label{eq:eta_par}
\end{align}
is the fraction by which the gas orbits slower than pure Keplerian rotation,
which is a function of the (local) gas scale height and pressure gradient
$\partial \ln P/\partial \ln a$.
Thus pebbles drift radially inwards within a short time-scale, 
\begin{align}
  t_{\rm d} \approx 5.5 \times 10^3
  \left( \frac{\tau_{\rm f}}{0.1} \right)^{-1} 
  \left( \frac{r}{10\,{\rm AU}} \right)
  {\rm \,yr}\,.
  \label{eq:drift-time}
\end{align}
A core embedded in the disc can accrete a sizable fraction of this radial
pebble flux, $f \approx 30$\,\% \citep[][see also Appendix\,\ref{sec:global}]{Morbidelli_2012}.

In this work, we use surface densities in pebbles comparable the MMSN
estimates, unless mentioned otherwise.
This approximation is supported by theoretical models of protoplanetary discs
that include dust growth by coagulation and radial drift of particles
\citep{Brauer_2008, Birnstiel_2012}.
A further discussion can be found in Appendix\,\ref{sec:global}.
This approach can also be justified observationally: measurements of the
spectral index of the dust opacity in protoplanetary discs reveal that a
significant fraction of solids grow to mm and cm sizes early on and remain
present over the life-time of the disc \citep{Ricci_2010}. 
In an accompanying paper \citep{Lambrechts_2014b} we further
study pebble accretion on global scales, including dust coagulation, pebble
drift and the growth of multiple cores. 
These results motivate the particle sizes and accretion rates used in this
study.

To conclude this section, we briefly highlight the main differences between
planetesimal and pebble accretion, which alter the accretion luminosity of the
core 
and therefore the critical core mass.
The pebble accretion rate given in Eq.\,(\ref{eq:Mdot_small_pebbles}) is the
maximal possible one, because in this regime one accretes from the full Hill
sphere, which is the largest possible gravitational reach of the core
\citep{Lambrechts_2012}.
The accretion of planetesimals, on the other hand, is significantly less
efficient compared to pebbles.
The planetesimal accretion rate can be expressed as fraction of the pebble
accretion rate,
\begin{align}
  \dot M_{\rm c,plan} \approx \psi \dot M_{\rm c,peb}.
  \label{eq:M_dot_planetesimals}
\end{align}
The efficiency of accretion, $\psi$, is equal to the ratio of the
core  radius $r_{\rm c}$ to Hill radius $r_{\rm H}$, 
\begin{align}
  \psi = \frac{r_{\rm c}}{r_{\rm H}}
    \approx 3\times10^{-4} \left( \frac{a}{\rm 10\,AU}\right)^{-1}
    \left( \frac{\rho_{\rm c}}{5.5{\rm g/cm}^3} \right)^{-1/3},
     \label{eq:p_eff}
\end{align}
where $\rho_{\rm c}$ is the material density of the core.
The reduced accretion rates are indicated by the grey lines in
Fig.\,\ref{fig:mvsmdotcrit}.
This result follows from the assumption that the planetesimal velocity
dispersion is equal to the Hill speed \citep{Dodson_2009,Dodson_2010}, $v_{\rm
H} = \Omega r_{\rm H}$, and gravitational focusing occurs from a radius $
(r_{\rm c}/r_{\rm H})^{1/2} r_{\rm H}$, which is smaller than the planetesimal
scale height $H_{\rm plan} = v_{\rm H}/\Omega = r_{\rm H}$. 
This leads to planetesimal accretion rates
\begin{align}
  \dot M_{\rm c,plan} \approx r_{\rm c} \Sigma_{\rm p} v_H, 
  \label{eq:M_dot_full}
\end{align}
with $\Sigma_{\rm p}$ now the surface density in planetesimals.
Collisional fragments of planetesimals ($0.1$-$1$\,km) have a reduced
scale height and can be accreted more rapidly by a factor
$1/\sqrt{\psi}$\,\citep{Rafikov_2004}.

Planetesimals are dynamically heated by the cores, which triggers a
fragmentation cascade. 
Because of efficient grinding of planetesimals to dust, the planetesimal
mass reservoir is reduced with time \citep{Kobayashi_2010, Kenyon_2008}. 
Therefore, core formation at 5\,AU requires massive planetesimal discs, at
least 10 times as massive as expected from the MMSN \citep{Kobayashi_2011}.
At wider orbital radii no significant growth occurs, although pressure bumps
caused by planet-triggered gap opening in the gas disc could increase the
accretion efficiency of fragments \citep{Kobayashi_2012}.
Global simulations furthermore highlight that growth by fragments is
inefficient, because they get trapped in mean motion resonances and push
planetary cores towards the star \citep{Levison_2010}. 
To overcome these issues, it has been proposed that fragmentation continues to
mm-cm sizes \citep{Ormel_2012, Chambers_2014}.
Pebbles, because of gas drag, do not suffer from destructive excitations or
resonant trapping.

\section{The growth of the proto-envelope}
\subsection{Accretion luminosity}
The attraction of the gaseous envelope is regulated by the growth of the
solid core.
Pebbles that rain down in the proto-atmosphere of the core deposit
their potential energy close to the core surface which provides the heat
necessary to support the envelope.
The luminosity of the planet is thus a simple function of the accretion
rate,
\begin{align}
  L = \beta G \frac{M_{\rm c} \dot M_{\rm c}}{r_{\rm c}},
  \label{eq:lum_acc}
\end{align}
where $r_{\rm c}$ is the radius if the core (for an extended discussion see
Appendix\,\ref{sec:Mcrit_calc}).
Depending on the composition of the accreted material, a fraction $1-\beta$ of
the mass attracted by the planet is lost by sublimation high up in the
atmosphere and pollutes the envelope with material of high molecular weight \citep{Hori_2011}. 
For this study we assume the bulk of the pebbles to be of cometary composition,
with a mass ratio of $\beta = 0.5$ of refractory elements to water ice.
In Appendix\,\ref{sec:composition}, we further discuss the influence of the
composition of the accreted material. 
Knowing the luminosity of the planet, we can now proceed to calculate the
structure of the hydrostatic envelope surrounding the core.

\subsection{The critical core mass}

There exists a \emph{critical} core mass where the inwards gravitational pull
of the core overcomes the  pressure support by the released accretion heat and
the envelope collapses.
We numerically investigate the envelope mass as function of the accretion
luminosity by constructing spherically symmetric envelopes in hydrostatic
equilibrium (Fig.\,\ref{fig:rhoT_0510} and \ref{fig:rhoT_2030}, further
description in Appendix\,\ref{sec:Mcrit_calc}).
We identify the critical core mass $M_{\rm c,crit}$ as the core mass for which
we no longer find a hydrostatic solution \citep{Mizuno_1980, Ikoma_2000}.
This occurs in practice when the mass bound in the envelope is comparable
to the mass of the core and the self-gravity of the gas atmosphere becomes
important \citep{Stevenson_1982}.
As a result the gas envelope falls onto the core on the Kelvin-Helmholtz
time-scale \citep[][see also Appendix\,\ref{sec:KH}]{Ikoma_2000, Piso_2013}. 
Earlier investigations show that higher accretion rates increase $M_{\rm
c,crit}$\,\citep{Ikoma_2000, Rafikov_2006}. 
In this study we have broadened the range of accretion rates studied to include
those from pebbles.
We also find that the critical core mass can be lowered significantly by
increasing the mean molecular weight through sublimation of icy material
in the deeper parts of the envelope where the temperature is above $T \approx
150$ K, in agreement with previous studies \citep{Stevenson_1982,Hori_2011}.
Figure\,\ref{fig:mvsmdotcrit} shows critical curves, in black, which
connect the critical core mass to the accretion rate (for a standard opacity
choice, see further Appendix\,\ref{sec:composition}).
We find that, unless accretion is interupted, pebble accretion leads to
critical core masses $\gtrsim 100$\,M$_{\rm E}$ between $5$-$30$\,AU, which is
an order of magnutude larger than the cores in the Solar System
\citep{Guillot_2005}.

\section{The pebble isolation mass}

\begin{figure}
  \centering
  \includegraphics[width=8.8cm]{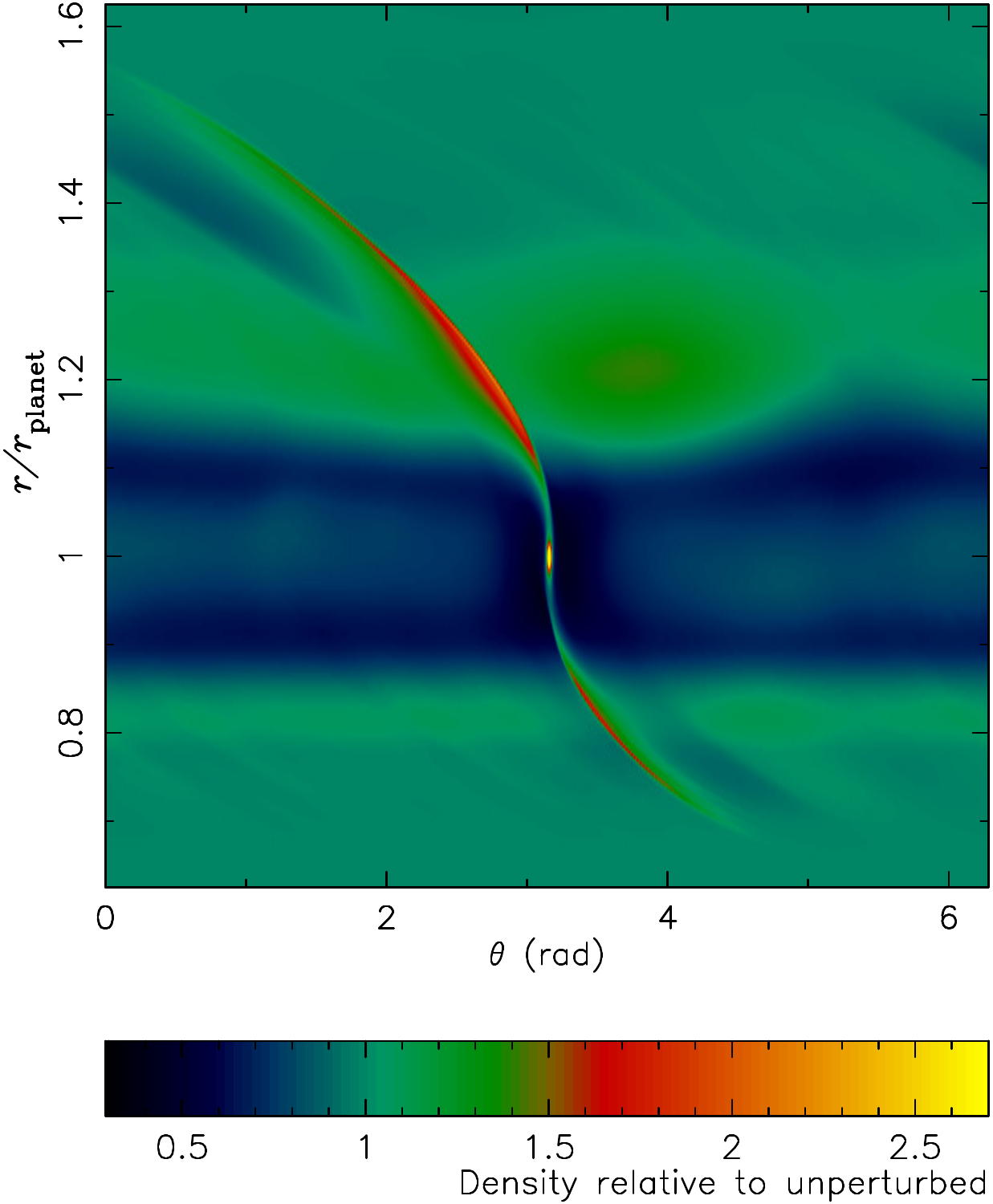}
  \caption{
  Gravitational perturbation of the midplane gas density from a $50$\,M$_{\rm
  E}$-planet embedded in a protoplanetary disc. 
  Displayed is a full annulus ($\theta=2\pi$-wide in azimuth) of the protoplanetary disc
  around the planet located at normalized radius $r=1$.
  The resulting pressure perturbation halts the radial migration of pebbles and
  thus the solid accretion onto the core. 
  The overdensities at radius $r=0.85$ and $r=1.2$ correspond to the regions
  with super-Keplerian rotation, that are further highlighted in
  Fig.\,\ref{fig:vtheta}. 
  A Rossby vortex, here centered at ($\theta=3.8,r=1.2$) has formed outside the
  orbit of the planet.
  }
  \label{fig:density_50Me}
\end{figure}

\begin{figure}
  \centering
  \includegraphics[width=8.8cm]{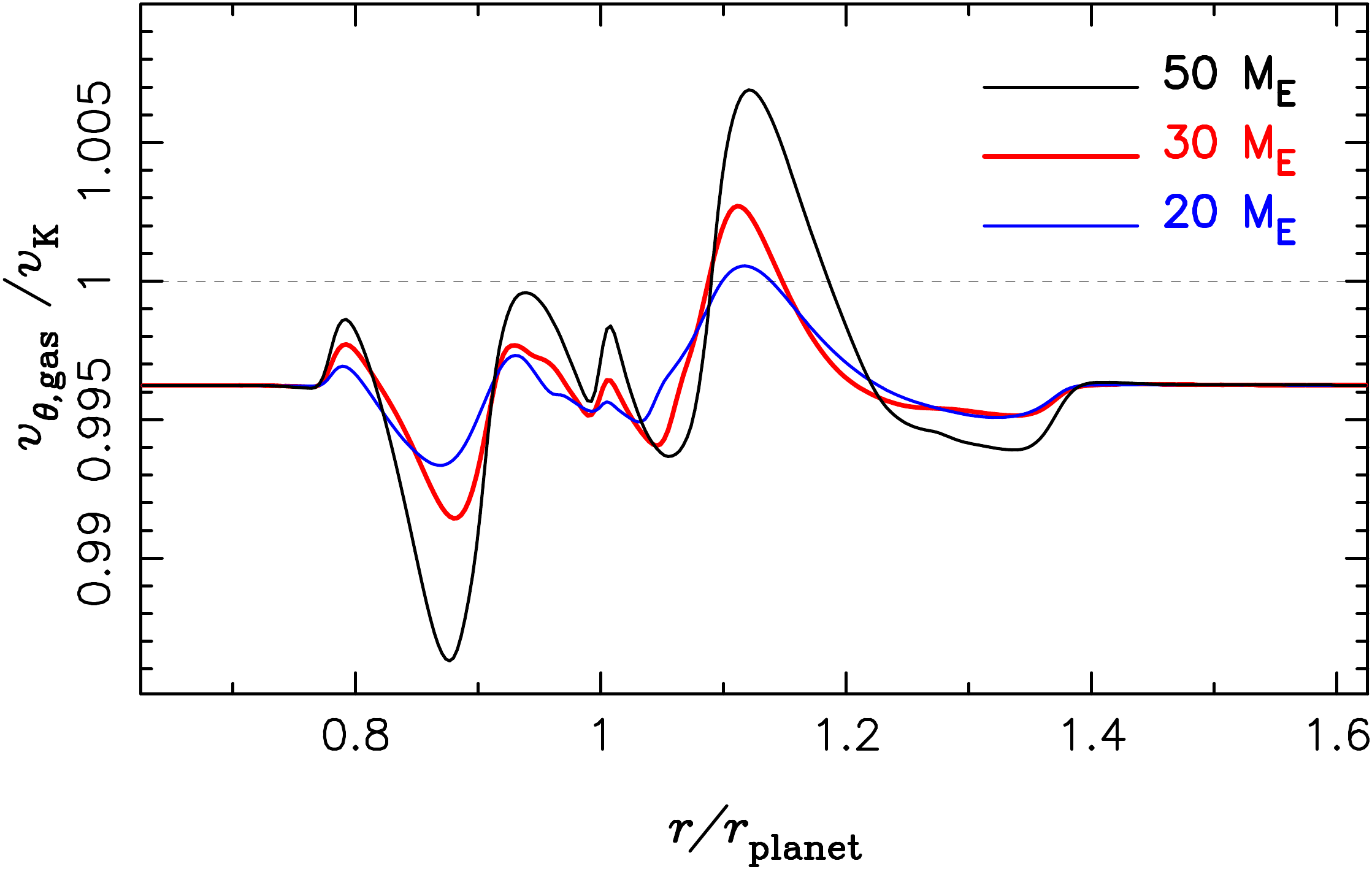}
  \caption{
  Deviation from the equilibrium sub-Keplerian velocity of the gas in the
  protoplanetary disc ($v_{\theta,{\rm gas}}$/$v_{\rm K}$), due to the presence
  of a planetary core located at a normalized radius of $r=1$ (here
  corresponding to an orbital radius of $a=5$\,AU). 
  For cores with masses above 20\,M$_{\rm E}$ (red curve) and higher
  (blue and black curves) the gas orbits faster than the Keplerian velocity
  (vertical dashed line) in a region outside the orbit of the planet. 
  This ring acts as a trap for pebbles drifting inwards and halts the accretion
  of pebbles onto the core.
  }
  \label{fig:vtheta}
\end{figure}

We now highlight the existence of a limiting mass for giant planets above which
no further pebbles are accreted.
Detailed 3-D numerical simulations of an annulus of the protoplanetary disc 
show that as the planet grows larger than the pebble isolation mass,
\begin{align}
    M_{\rm iso} \approx 20 \left( \frac{a}{5{\rm \,AU}}\right)^{3/4}
    {\rm M}_{\rm E},
    \label{eq:isomass} 
\end{align}
local changes in the pressure gradient modify the rotation velocity of the gas, 
which halts the drift of pebbles to the core
(Eq.\,\ref{eq:rad_vel}-\ref{eq:eta_par} and
Fig.\,\ref{fig:density_50Me}-\ref{fig:vtheta}).
The value of the pebble isolation mass depends dominantly on the orbital radius
through the disc aspect ratio, $M_{\rm iso} \propto (H/a)^3$, see further
Sec.\,\ref{sec:FARGO}. 
Therefore pebble isolation becomes harder to attain at wider orbital
separations in flaring discs.

\subsection{Calculation of the pebble isolation mass}
\label{sec:FARGO}

We now determine at which mass a planet can isolate itself
from the flux of pebbles, and the dependence of the result on the scale height
of the disc and the planet's location.
A planet can perturb the gas-disc enough that the latter can become
super-keplerian over a narrow ring just outside the orbit of the planet
\citep{Paardekooper_2006,Morbidelli_2012}. 
This happens when the planet is capable
of perturbing significantly the density of the disc, changing locally the sign
of the radial density gradient. In the ring where the gas is super-keplerian,
the action of gas-drag reverses. Pebbles are pushed outwards, instead of
inwards. Thus, the pebbles have to accumulate at the outer edge of this ring,
instead of migrating all the way through the orbit of the planet. The
accretion of pebbles by the planet now suddenly stops.
Possibly, the enhancement in pebble density at the edge of the ring can lead to
the formation of new large planetesimals and even of new cores
\citep{Lyra_2008, Kobayashi_2012}.

An isolation mass of about 50\,M$_{\rm E}$, for a disc with a scale height of
5\%, was previously suggested in \citet{Morbidelli_2012}.
However, the authors used 2-dimensional
simulations, which forced them to use a large smoothing parameter (equal to 60\%
of the planet's Hill radius) in the planet's gravitational potential. The use
of a large smoothing parameter weakens the gravitational perturbations of the
planet on the disc, so that their estimate is probably an overestimate.

To overcome this problem, we used here a new 3D version of the code
FARGO \citep{Masset_2000,Lega_2013}. The new code also
handles the diffusion of energy and stellar irradiation, but we used here its
isothermal version for simplicity. The 3D code adopts a cubic approximation for
the gravitational potential of the planet \citep{Kley_2009}, which is
somewhat equivalent to assuming a very small smoothing parameter in a standard
potential.

We modeled a disc from $0.625$ to $1.62$ in radius, the unit of distance being
the radius of the planet's orbit, with an aspect ratio of 5\% and a viscosity
given by an $\alpha$ prescription \citep{Shakura_1973} with
$\alpha_{\rm t}=6\times
10^{-3}$. The radial boundary conditions were evanescent, which prevented
reflection of the spiral density wave. The boundary condition in co-latitude
was instead reflecting. The resolution was $320\times720\times32$ in the
radial, azimuthal and co-latitudinal directions, respectively. 
We did a simulation with a planet of 20, 30 and 50\,M$_{\rm E}$.
The simulations have been run for 60 orbits, when the disc seemed to have
reached a stationary structure.
For the purpose of identifying the pebble isolation mass, it is not necessary
to explicitly model the trajectories of particles between $\tau_{\rm f} =
0.001-1$ \citep{Morbidelli_2012}. 
We therefore limit ourselves to calculating the pressure perturbation able to
halt the radial particle drift (Eq.\,\ref{eq:rad_vel}).

Figure\,\ref{fig:vtheta} shows the ratio between the azimuthal
velocity of the disc and the Keplerian velocity as a function of radius.
For each radius, the azimuthal velocity has been computed on the
mid-plane of the disc; its average over the azimuth has been mass-weighted. We
also computed the vertically averaged azimuthal velocity (all averages were
mass-weighted) and found essentially the same radial profile.  As one can see
in the figure, away from the planet, the disc is uniformly sub-keplerian (the
azimuthal velocity is 0.9962 times the Kepler velocity). Instead, the planet
induces strong perturbations in the gas azimuthal velocity in its vicinity. 
In particular for the case of the 30\,M$_{\rm E}$ core, there are two strong
signatures, associated with the edges of the shallow gap that the planet opens
in the disc: a dip at at $r=0.88$ where the gas is strongly sub-Keplerian, and
a peak at $r=1.11$ where the gas exceeds the Keplerian velocity (the azimuthal
velocity is 1.0025 times the Kepler velocity). 
In this situation, the pebbles are expected to stop drifting at $r\approx 1.15$,
where the gas turns from sub-Keplerian (beyond this distance) to
super-Keplerian (inside this distance).

Performing a simulation with a 50\,M$_{\rm E}$ planet we checked that, as
expected, the velocity perturbation is linear in the mass of the planet. Thus,
the planet-mass threshold for turning the disc barely super-Keplerian is
$M_{\rm iso} \approx 20\,{\rm\,M}_{\rm E}$.
We verified this result with a simulation with a planet of this mass.

We also checked, with a simulation with a 5 times smaller value of the
turbulent $\alpha_{\rm t}$ parameter, that the azimuthal velocity has a
negligible dependence on the viscosity of the disc.
This was expected because for a disc undergoing perturbations by a small
planet, the resulting disc structure is dominated by disc's internal pressure
and hence its aspect ratio \citep{Crida_2006}.
In completely inviscid discs, the estimate for the isolation mass is likely
smaller by no more than a factor $\approx 2$ \citep{Zhu_2014}.

The dependence of $M_{\rm iso}$ on the aspect ratio can be estimated            analytically.
In fact, in the limit of negligible viscosity, scaling the disc's aspect ratio
$H/a$ proportionally to the normalized Hill radius of the planet $r_{\rm
H}/a$, and adopting $r_{\rm H}$ as basic unit of length, the equations
of motion for the fluid become independent of the planet mass
\citep{Korycansky_1996}.  Given that the perturbation in azimuthal velocity is
linear in the planet's mass \citep{Korycansky_1996}, the result implies that the
perturbation in azimuthal velocity has to be proportional to $(H/a)^3$.
With this result, we can now conclude with the dependency of $M_{\rm iso}$ on
the location of the planet in a given disc. 
Assuming that a disc is flared like the MMSN, as 
$H/a = 0.05 (a/5\,{\rm AU})^{1/4}$, we obtain the result expressed in Eq.\,(\ref{eq:isomass}).
Alternatively, for discs irradiated by the star the gas scale height goes as
$H/a \propto a^{2/7}$ \citep{Chiang_1997}. 
The exact value of $H/a$ depends on the level of viscous heating in the inner
disc \citep{Bitsch_2013}, but for moderate disc accretion rates is similar to
the MMSN estimate at $5$\,AU.
Therefore, the scaling for the isolation mass in an irradiated disc takes the
form $M_{\rm iso} = 20\,{\rm M}_{\rm E} (a/5\,{\rm AU})^{8/7}$, which is
slightly steeper than the MMSN-estimate.

\subsection{Implications of a pebble isolation mass}
The existence of this pebble isolation mass has three major implications
that show the advantage of pebble accretion over planetesimal
accretion in setting the conditions for envelope collapse.
Firstly, when a giant planet grows to a mass beyond the pebble isolation mass,
solid accretion will be \emph{abruptly} terminated.
The accretion luminosity is quenched and the critical core mass drops to a
value much smaller than the pebble isolation mass
(Fig.\,\ref{fig:mvsmdotcrit}), which triggers a phase of rapid gas accretion.
This is in sharp contrast with core growth by planetesimal accretion, where the
continuous delivery of solid material delays the gravitational collapse by
millions of years \citep{Pollack_1996}. 
Halting planetesimal accretion to overcome this difficulty has been
previously proposed \citep{Hubickyj_2005}, but the formation of a clean gap in a
planetesimal disc demands small planetesimals with a low surface density
\citep{Shiraishi_2008}, which is inconsistent with models of core growth with
planetesimals \citep{Levison_2010,Kobayashi_2011}.

Secondly, the low value of the pebble isolation mass resolves an apparent
paradox faced by any growth scenario for giant planets:
the high accretion rates necessary to form cores before gas dissipation results
in critical core masses that are too large by an order of magnitude
($\sim$$100$\,M$_{\rm E}$, Fig.\,\ref{fig:mvsmdotcrit}). Fortunately, the
self-shielding nature of pebble accretion yields much lower core masses in
5--10 AU orbits.

Thirdly, the pebble isolation mass introduces a natural sharp divide of the
giant planets into two classes: gas giants and ice giants. 
The latter category are those cores that did \emph{not} reach the pebble
isolation mass before disc dissipation. 
Therefore these planets never stopped accreting pebbles during the life-time of
the gas disc and the resulting accretion heat prevents unpolluted H/He
envelopes from becoming unstable and undergoing runaway gas accretion.
As a result, low-mass cores only contract low-mass envelopes with increased
mean molecular weight (Fig.\,\ref{fig:mvsmdotcrit}), which occurs after pebbles
sublimate below the ice line and the released water vapour is homogeneously
distributed by convection.
Formation of ice giants in this model is thus different from the classical core
accretion scenario, where ice giants are giant planets that had their growth
prematurely terminated by demanding gas dispersal during envelope contraction
\citep{Pollack_1996,Dodson_2010}.

\section{Pebble accretion in the Solar System}

\begin{figure}
  \centering
  \includegraphics{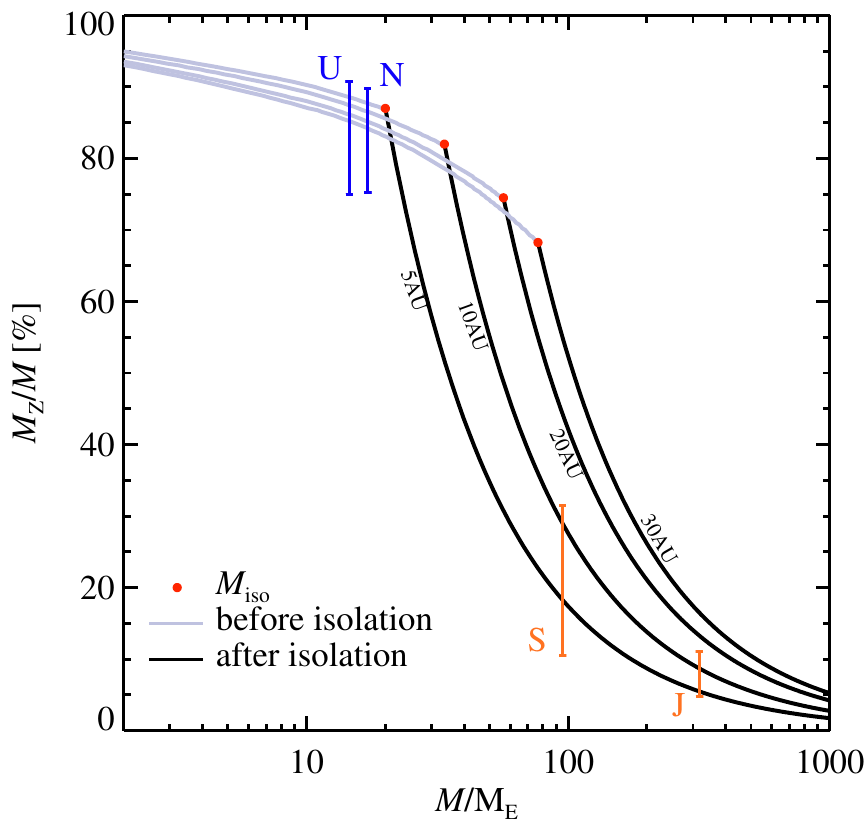}
  \caption{
  Total heavy element mass fraction as function of the total mass $M$ of the
  giant planet, at different orbital radii (5,10,20,30\,AU). 
  Planets that do not grow beyond the pebble isolation mass 
  (red dots)
  remain core-dominated, while those that grow larger will have most of their
  mass in gas.
  Estimates of the composition of Uranus and Neptune 
  \citep[blue error bars,][]{Helled_2011} 
  agree well with the prediction made in this paper for planets formed in the
  outer disc.
  Similarly, for the gas giant planets Jupiter and Saturn we find a good
  agreement between the $5$-$10$\,AU curves and the total heavy element mass
  estimated by \citet{Guillot_2005}, indicated by the orange errorbars.
  In order to create this figure, we numerically calculated
  the composition of the planet when it becomes critical, taking into account
  the pollution of the envelope, for planet masses below the pebble isolation
  mass (light blue curves).  
  When the planet reached a mass larger than
  the pebble isolation mass, only nebular gas was added for further growth
  (black curves).
  Here we present results with refractory fraction $\beta=0.5$, but results
  depend  weakly on the choice of $\beta$ between $0.1$-$1$.
  }
  \label{fig:ssfit}
\end{figure}

The pebble isolation mass increases with orbital radius (Eq.\,\ref{eq:isomass},
Fig.\,\ref{fig:mvsmdotcrit}), so the ice giants can not have formed too close
to the host star ($<$$5$\,AU) where they would have become gas giant planets.
Such close formation distances for the ice giants are anyway not favoured in
current models of the early migration history of the Solar
System \citep{Walsh_2011}:
following the gas giants, the ice giants migrated inwards, but remained outside
$5$\,AU, and subsequently moved outwards to distances between \mbox{$11$--$17$}\,AU which
are preferred to explain the late time orbital evolution after disc
dissipation \citep{Tsiganis_2005}.
The formation of the ice giants in our model is therefore compatible
with the understanding of planetary migration in the Solar System.
One intriguing option that was previously not possible, is the
approximately \emph{in situ} formation of ice giants beyond $20$\,AU, since pebble accretion is sufficiently fast.

\subsection{Planetary composition}

By combining our calculations of the pebble isolation mass and
the critical core mass (including the effect of pebble sublimation) we can
calculate the heavy element mass fraction as function of the total planet mass
(full description in Appendix\,\ref{sec:critmasscalc}).
We do not compare our results directly with the inferred core masses of the
giant planets in the Solar System, but instead with the total heavy element
mass, as it is better constrained and transport could have
occurred from core to envelope \citep{Guillot_2005}.
We find good agreement for Jupiter and Saturn between $5$--$10$\,AU, while
Uranus and Neptune could have formed at similar or wider orbits, as
illustrated in Fig.\,\ref{fig:ssfit}.

In contrast, the composition of the giant planets is difficult to reproduce with
planetesimal accretion.
For ice giants, planetesimal accretion is too slow at wide orbits, which would
make the critical core mass too low (Fig.\,\ref{fig:mvsmdotcrit}). 
For gas giants, late-time planetesimal accretion after runaway gas
accretion can not add the significant mass fraction of heavy elements in their
envelopes, because planetesimal capture rates are small \citep{Guillot_2000}.

\begin{figure}
  \centering
  \includegraphics{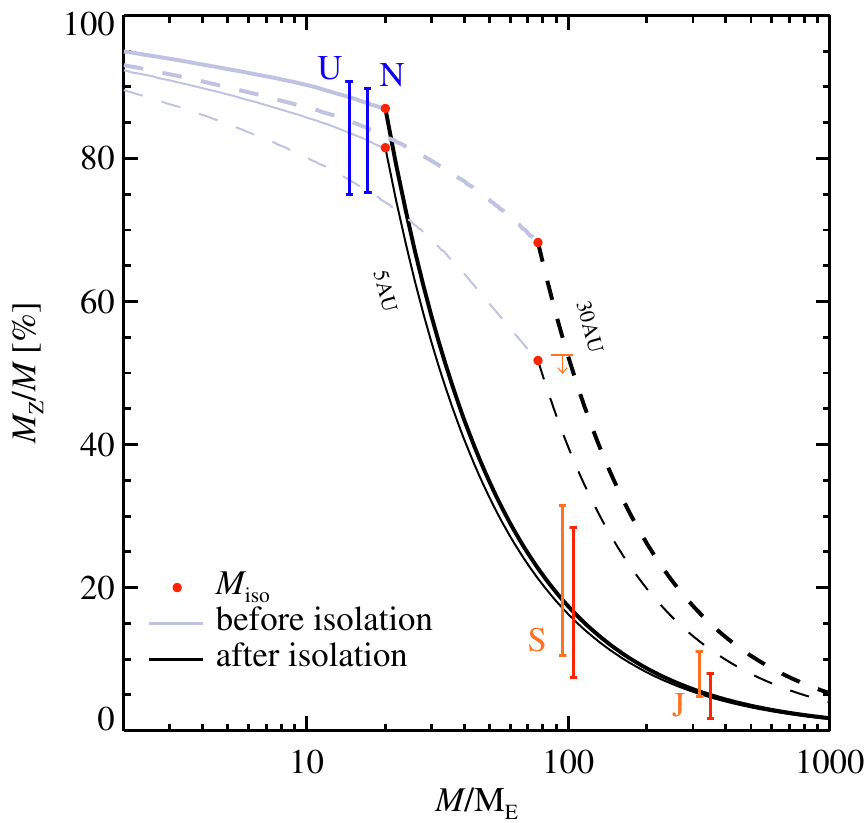}
  \caption{
  Similar to Fig.\,\ref{fig:ssfit}, but representing planetary
  compositions of planets located at $5$ and $30$\,AU (full and dashed
  lines respectively) for different values of the accretion rate
  efficiency $\epsilon$. 
  The thick solid lines correspond to the case where the full pebble accretion
  luminosity is released ($\epsilon=1$) and the lower thin lines of the same
  color to a 10 times smaller luminosity ($\epsilon=0.1$). 
  The red error bars, slightly displaced to the left for readability, give the
  heavy element mass, without the contribution of heavy elements added by late
  time gas accretion in a disc with increased dust-to-gas ratio
  \citep{Guillot_2006}, as suggested to explain the noble gas abundances in
  Jupiter.
  The orange downwards arrow is an upper limit on the heavy element content of
  Saturn if layered convection occurs in the envelope \citep{Leconte_2013}. 
  }
  \label{fig:ssfit_1030}
\end{figure}

Our encouraging correspondence between the model and the composition of
the giant planets does not depend strongly on the assumptions made on the
composition or surface density of pebbles.
For Fig.\,\ref{fig:ssfit_1030} we have repeated our analysis, but with a
$10$\,times lower accretion luminosity,
by introducing a fudge factor $\epsilon=0.1$ into Eq.\,(\ref{eq:lum_acc}).
Such a reduced luminosity could for example be the result of the bulk
composition of the pebbles to be largely in ice, reducing the refractory
fraction $\beta$ by a factor $\epsilon$. 
Or, alternatively, the consequence of a lower surface density in pebbles
compared to an MMSN-estimate, which would lower
the accretion accretion rate and thus the luminosity by the same factor.
These lower efficiencies do not change the fit for the solar system gas giants,
while ice giants are in fact better matched, if they formed closer
towards the star near the ice line, compared to  their current orbits.

We briefly address two caveats concerning the measured heavy element
masses of the gas giants.
Recently, it has been suggested that layered convection is important for
Saturn \citep{Leconte_2013}, which allows for a higher total heavy element mass
(an upper limit on the heavy element content is indicated by the downwards
arrow). 
Also it is possible that a fraction of the heavy elements gets delivered
through gas accretion at a time close to the dissipation of the protoplanetary
disc, which would explain the noble gas abundances of Jupiter. In
Figure\,\ref{fig:ssfit_1030}, the red error bars are corrected for the heavy
elements accreted in this stage after pebble accretion.

\section{Exoplanets}

Our proposed formation model for giant planets is applicable to extrasolar
planetary systems as well.
If a core reaches the pebble isolation mass before disc dissipation, it becomes a gas giant. 
However, outside an orbital radius $a_{\rm lim}$ the growth of cores will be
too slow and isolation masses too high for pebble isolation to occur within the
disc life-time $\tau_{\rm disc}$. Those cores end up as ice giants.
In a disc with a solid surface density equal to the MMSN ($\epsilon=1$),
we find the limiting semi-major axis for gas giants to be
\begin{align}
  a_{\rm lim} \approx 95\,
  \epsilon^{4/5}
  \left( \frac{\tau_{\rm disc}}{2{\rm \,Myr}} \right)^{4/5}
  {\rm \,AU}\,.
  \label{eq:rl_bis}
\end{align}
In massive protoplanetary discs, $\epsilon$ is near unity or larger
and $a_{\rm lim}$ is thus located far out in the disc. 
In such a case, wide-orbit gas giants, such as the four planets around the A
star HR8799 located at $a = 15$--$70$\,AU\,\citep{Marois_2010}, can form in
situ. 
We predict that their cores will be large ($\sim$$50$--$100$\,M$_{\rm E}$),
because of the high pebble isolation mass at large orbital distances (or from
the similarly large critical core masses for pure H/He envelopes as seen in Fig.\,\ref{fig:mvsmdotcrit}).
Such solid-enriched compositions are supported by models of gas-giant
exoplanets now in close orbits (but which likely formed at wider orbits) that
show total heavy element masses $\sim$$100$\,M$_{\rm E}$, much larger than
planetesimal isolation masses in classical core accretion \citep{Pollack_1996}. 
For example, the gas giant CoRoT-10 b has mass of 870\,M$_{\rm E}$, of which
approximately 180\,M$_{\rm E}$ are in heavy elements \citep{Bonomo_2010}.
Similarly, \mbox{Corot-13 b}, \mbox{14 b}, \mbox{17 b}, \mbox{20 b} and
\mbox{23 b} all have $\gtrsim$$70$\,M$_{\rm E}$ of heavy elements and are
significantly enriched with respect to the host star
metallicity \citep{Moutou_2013}.

Nevertheless, more commonly giant planets at wide orbits will be ice giants. 
Indeed relaxing the efficiency somewhat results in, for example, $a_{\rm lim} =
15$\,AU for $\epsilon=0.1$ (in good agreement with the Solar System). 
This seems also to be broadly consistent with the low occurrence of gas
giants at very wide wide orbits inferred from direct imaging surveys
\citep[the fraction of FGKM stars with planetary companions $\gtrsim 2$ Jupiter masses beyond 25\,AU is below 20\%, ][]{Lafreniere_2007}.
Altogether, pebble-driven envelope attraction predicts an orbital and
compositional dichotomy, similar to the solar system, between gas giants and
ice giants in extrasolar systems.

\section{Summary}

The model we propose can be summarised as follows. 
First, rocky/icy particles grow throughout the disc by sticking
collisions and condensation around ice lines. 
When solids reach mm-cm sizes they start to decouple from the gas in the
protoplanetary disc and drift towards the sun. 
Then, hydrodynamical concentration mechanisms of pebbles, such as the streaming
instability or vortices, lead to the formation of a first generation of planetesimals.  
Subsequently, the largest of these planetesimals continue to grow by accreting
from the flux of pebbles drifting through the disc.
The high accretion rates result in core formation on a short timescale, within
the disc life-time. 
Such fast growth implies that the atmospheres around the growing cores are
strongly supported by the accretion heat. 

We show that the evolution of the gas envelope is different depending on
whether the core reaches the pebble isolation mass (or not), resulting in
respectively gas giants or ice giants.
When a core grows sufficiently large, around $20$\,M$_{\rm E}$ at 5\,AU, it can
halt accretion of solids onto the core by gravitationally perturbing the
surrounding gas disc. This creates a pressure bump that traps incoming
pebbles. 
When the core gets isolated from pebbles, the envelope is no longer
hydrostatically supported by accretion heat, and gas can be accreted in a
runaway fashion. 
This leads to the formation of gas giants like Jupiter and Saturn.

Cores in wider orbits need to grow more massive than $20$\,M$_{\rm E}$ to
reach isolation, because of the steep increase in the gas scale height in
flaring discs. 
Therefore, wide-orbit cores that do not grow larger than $50$-$100$\,M$_{\rm
E}$ during the gas disc phase remain supported by accretion heat and in hydrostatic balance. 
This offers an explanation for the occurrence of planets that only attract a
thin envelope of hydrogen and helium, mixed with large amounts of water vapour
released by sublimation of icy pebbles, such as the ice giants Uranus and
Neptune in our Solar System.

We demonstrate that this single model explains the bulk
composition of \emph{all} the giant planets in the Solar System.
Additionally, the pebble accretion scenario can be tested by studying
exoplanet systems. 
We find that most exoplanets in wide orbits will be similar to ice giants.
Only when cores grow very large ($\gtrsim 50$\,M$_{\rm E}$), within the disc
life-time, can gas giants form in wide orbits. 
We therefore predict a high solid enrichment for gas-giants exoplanets in wide
orbits, like those found in the HR8799 planetary system.

\begin{acknowledgements}
M.L. thanks K. Ros, B. Bitsch and T. Guillot for comments. 
The authors also wish to thank the referee for thoughtful feedback which
helped improve the manuscript.
A.J.\,and M.L.\,are grateful for the financial support from the Royal Swedish
Academy of Sciences and the Knut and Alice Wallenberg Foundation.
A.J.\,was also acknowledges funding from the Swedish Research Council (grant
2010-3710) and the European Research Council (ERC Starting Grant
278675-PEBBLE2PLANET).
A.M. thanks the French ANR for support on the MOJO project.

\end{acknowledgements}

\bibliographystyle{aa}        
\bibliography{references_sup} 

\appendix

\section{Pebble surface density and accretion efficiency}
\label{sec:global}
The pebble surface densities inferred from the Minimum Mass Solar Nebula, used
throughout this paper, are broadly consistent with surface densities calculated
in more detailed models of protoplanetary discs that combine dust growth and
the drift of pebbles. 
The reconstruction of the MMSN is based on the questionable assumption that
planets grow in situ out of all the material which is available locally. In
contrast, observed protoplanetary discs are larger and show temporal evolution
in both gas and dust components. This evolution is understood to be the
consequence of a drift-limited dust growth by coagulation \citep{Brauer_2008}. 
As demonstrated by \citet{Birnstiel_2012}, the dominant particle size is well
characterized by an equilibrium between the local growth time scale and the
drift timescale and lies in a narrow size range between $\tau_{\rm f}=0.01-0.1$
at distances between $5$-$30$\,AU. 
The surface density of these pebbles also remains high during the disc
life-time. 
During the first $\approx$\,$1$\,Myr the initial dust-to-gas ratio of 0.01 is
maintained, only to decay to $\approx$\,$0.001$ after $\approx$\,$3$\,Myr.
Therefore the employed surface densities in the paper are adequate, certainly
given the short timescales on which cores grow by pebble accretion. 
Furthermore, we also demonstrate our model holds for surface densities reduced
by an order of magnitude (Fig.\,\ref{fig:ssfit_1030}).

The total pebble mass in the disc needed for our model is also consistent
with protoplanetary disc observations. 
Pebbles from the drift equilibrium model are efficiently accreted at a rate 
\begin{align}
  \dot M_{\rm c} = 2 \left(\frac{\tau_{\rm f}}{0.1}\right)^{2/3} r_{\rm H}
  \Sigma_{\rm p} v_H 
  \label{eq:Mdoti_small}
\end{align}
as demonstrated by numerical simulations for particles in the
$\tau_{\rm f}=0.01$-$0.1$ range \citep{Lambrechts_2012}.  
The radial flux of pebbles through the disc is given by
\begin{align}
  \dot M_{\rm drift}  &= 2 \pi a \Sigma_{\rm p} v_r 
  \approx 4 \pi \Sigma_{\rm d} a \tau_{\rm f}  \eta v_K,
  \label{eq:M_drift}
\end{align}
where $v_{\rm K}$ is the Keplerian velocity at orbital radius $a$ and $\eta$ is
a dimensionless measure of radial gas pressure support (Eq.\,\ref{eq:eta_par}).
The embryo will accrete the fraction $f=\dot M_{\rm c}/\dot M_{\rm
drift}$ of these solids, 
\begin{align}
  f &\approx \frac{20}{4 \pi}  \eta^{-1} 
  \left( \frac{\tau_{\rm f}}{0.1} \right)^{-1/3} 
  \left( \frac{r_H}{a} \right)^{2/3}\\
  &= 0.35 \left( \frac{\tau_{\rm f}}{0.1} \right)^{-1/3}
      \left( \frac{M_{\rm c}}{20\,{\rm M}_{\rm E}} \right)^{2/3}
      \left( \frac{a}{\rm 5 AU} \right)^{-1/2}.
  \label{eq:filter}
\end{align}
The filtering factor $f$ itself does not explicitly depend on $\Sigma_{\rm d}$,
but does depend on $\eta$, thus in regions with reduced pressure support
(pressure bumps) the efficiency could be higher.
The necessary mass in pebbles in the disc can be estimated from $dM_{\rm p} =
f^{-1} dM_{\rm c}$,
\begin{align}
  M_{\rm p} \approx 165\,{\rm M}_{\rm E} 
  \left( \frac{M_{\rm c}}{20\,M_{\rm E}} \right)^{1/3}
         \left( \frac{a}{\rm 5 AU} \right)^{1/2}
         \left( \frac{\tau_{\rm f}}{0.1} \right)^{1/3},
  \label{eq:M_cum}
\end{align}
in order to grow the core to 20\,M$_{\rm E}$ (starting from $0.1$\,M$_{\rm E}$, but the integral only weakly depends on this choice).
Therefore the mass reservoir in the outer disc is on this order, and the total
disc mass should be about 0.05 solar mass (for a standard $Z=0.01$
metallicity).
Also, because of the low value of $f$ for small core masses, no larger disc
masses are needed to form more cores in the disc.

The disc masses needed for our model are consistent with observations. The best studied
protoplanetary disc, the disc of the star TW Hydrae, has a gas mass of
0.05\,M$_\odot$ \citep{Bergin_2013} and the mm and gas distribution are well
described by the model of \citet{Birnstiel_2012}. 
From mm-surveys it seems such disc masses for solar like stars show a large
spread between $10^{-4}$--$10^{-1}$\,M$_\odot$ \citep{Andrews_2013} and
therefore the disc mas in our model lies in the higher end of this
distribution. However, disc mass estimates are based on an assumed ratio of gas
to mm-sized dust of 100, and therefore these mass estimates may be lower limits. 

The above analysis does not take into account the presence of ice lines. In
these regions particle sizes and the local surface density are set by a
condensation-sublimation cycle across the ice line, resembling hail formation
\citep{Ros_2013}. This is very different from the coagulation-drift equilibrium
situation discussed above.
Likely, around the various ice lines in protoplantary discs (H$_2$O, CO$_2$,
CO) solid surface densities in pebbles of rather large size
($\tau_{\rm f}$$\sim$$0.1$) greatly exceed MMSN estimates, promoting fast core growth.
Furthermore, recondensation of sublimated pebbles onto particles exterior of the
ice line reduces the loss of ices.

A more thorough discussion of core growth can be found in
\citet{Lambrechts_2014b}, where we investigate embryo growth in a global model
that includes dust growth and the drift of pebbles.

\section{Calculating the critical core mass}
\label{sec:Mcrit_calc}

\subsection{Structure of the proto-envelope}
\label{sec:method}

\begin{figure*}
  \centering
  \includegraphics{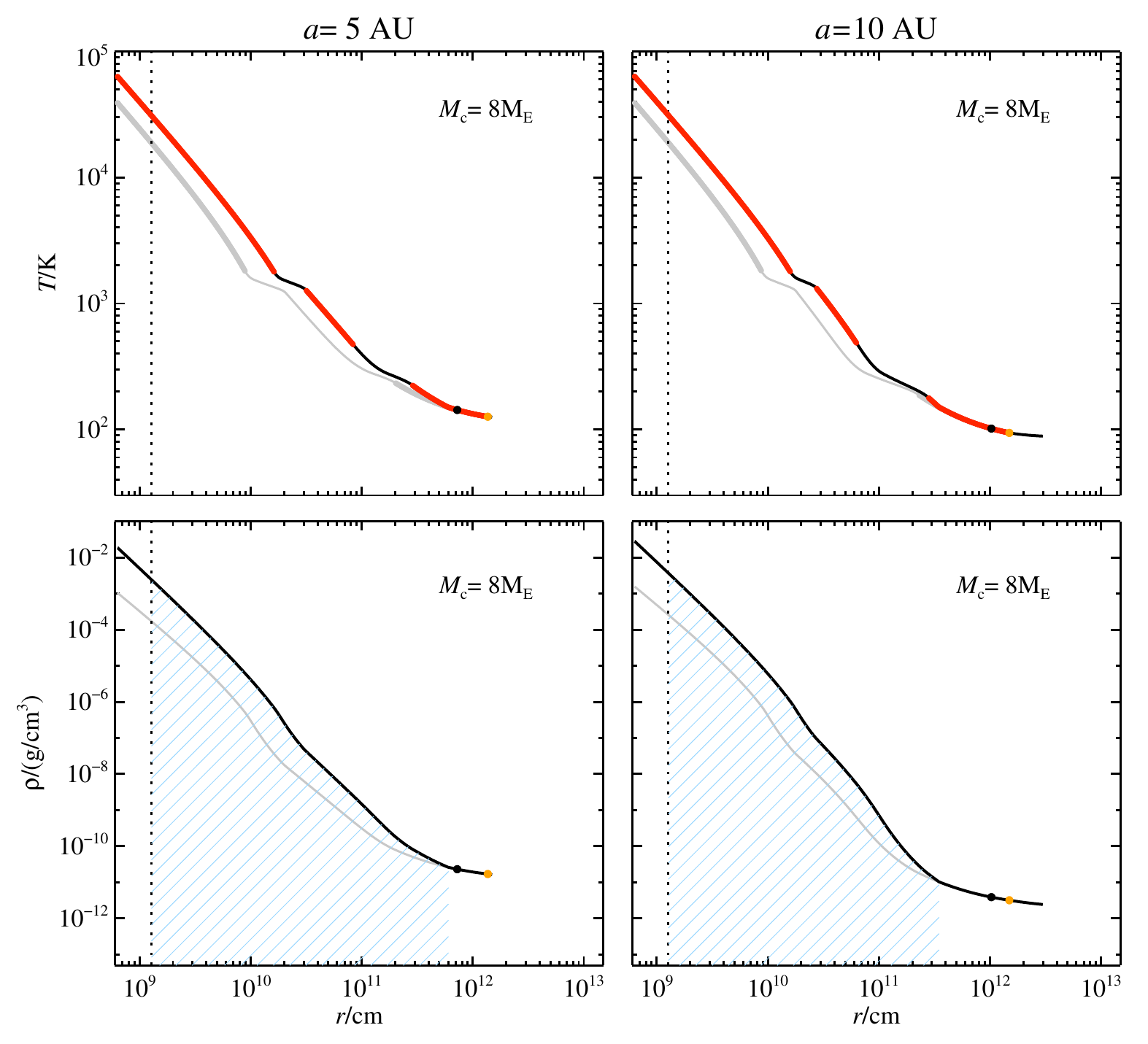}
  \caption{
    Dependence of the hydrostatic envelope on orbital radius and envelope
    pollution. The left panels give the temperature (top) and density profile
    (bottom), for both a $50\%$ polluted and unpolluted atmosphere (grey curve)
    of an 8\,M$_{\rm E}$ core accreting pebbles at $5$\,AU.  Thick lines
    represent the regions where heat transport by convection dominates. The
    profile is given starting from the connection point to the Hill sphere. 
    The yellow circle gives the photosphere of the envelope, while the black
    circle indicates the Bondi radius (the distance where the escape speed
    equals the local sound speed). 
    The depth at which the envelope density is enhanced by water vapour is
    marked by the dashed blue lines in the lower panels. 
    The dotted line represents the  core radius (assuming $\rho_{\rm
    c}=5.5$\,g/cm$^3$).  The right column is similar, but for a planet orbiting
    at 10\,AU.
  }
    \label{fig:rhoT_0510}
\end{figure*}

\begin{figure*}
  \centering
  \includegraphics{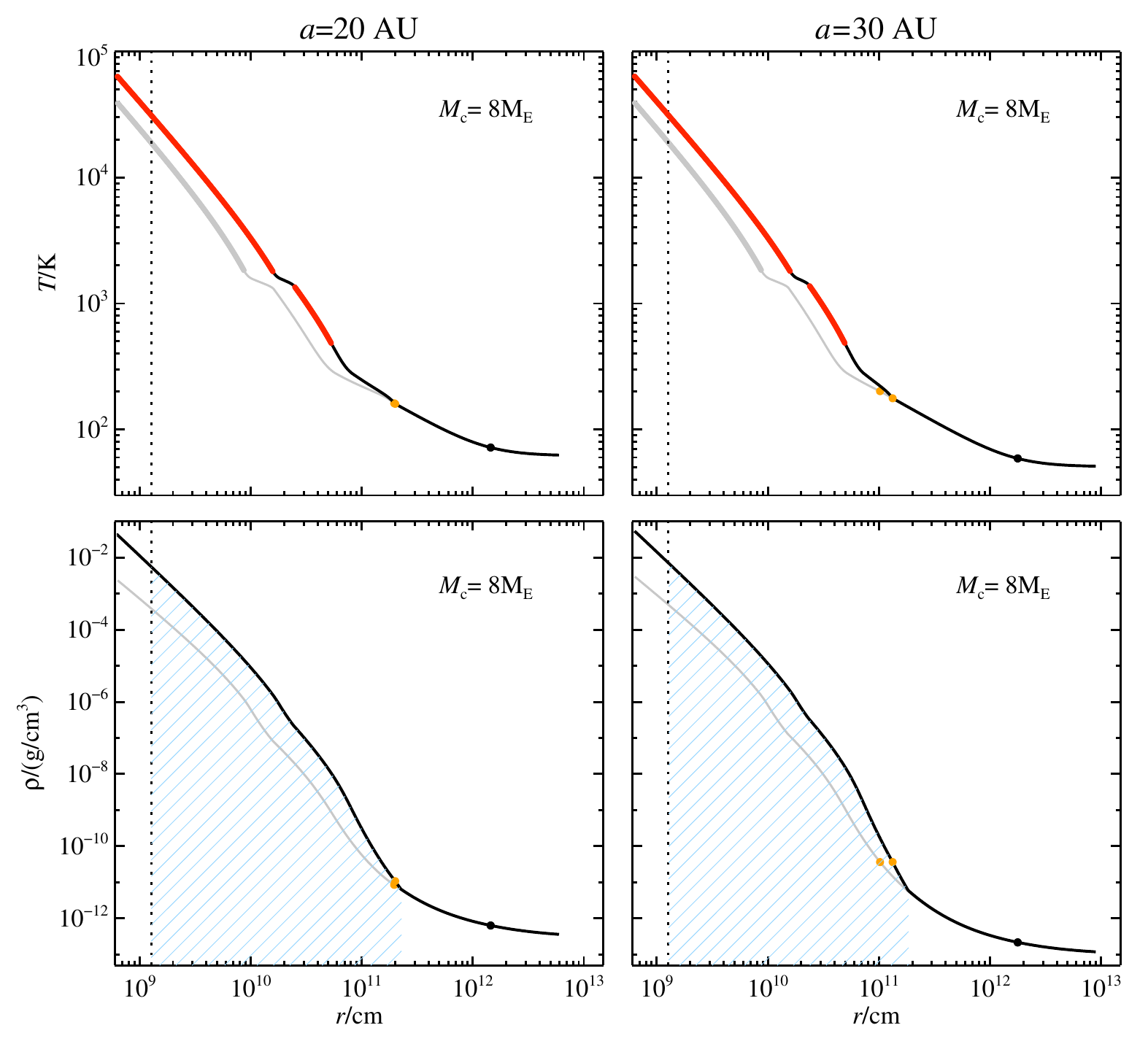}
  \caption{
  Comparison of the structure between the hydrostatic envelope enriched by
  H$_2$O steam and a pure H/He atmospheres, around embryo cores
  of $8$\,M$_{\rm E}$ located at $20$ and $30$\,AU. 
  Labels are similar to Fig.\,\ref{fig:rhoT_0510}. 
  }
  \label{fig:rhoT_2030}
\end{figure*}

We numerically solve the standard equations for planetary atmospheres
\citep{Kippenhahn_1990}.
The envelope is assumed to be spherically symmetric and in hydrostatic balance,
\begin{align}
  \frac{dP}{dr} = -\frac{GM(r)\rho}{r^2}.
  \label{eq:hydrostat}
\end{align}
Here $G$ is the gravitational constant, $P$ the pressure and $\rho$ the density
at position $r$ from the centre of the planet.
The mass interior to a radius $r$ is given by $M(r)$.
Mass continuity is guaranteed by
\begin{align}
  \frac{dM(r)}{dr} = 4 \pi r^2 \rho.
  \label{eq:atmo_cont}
\end{align}
Energy can be transported either by radiation diffusion or convection in
optically thick regions. 
Convective heat transport is triggered when
\begin{align}
   \frac{\partial \ln T}{\partial \ln P} > \frac{\gamma-1}{\gamma},
   \label{eq:conv_vs_rad}
\end{align}
where T is the local temperature and $\gamma$ is the adiabatic index. 
Convective transport takes the form
\begin{align}
 \frac{dT}{dr}  
  = - \frac{\gamma-1}{\gamma} \frac{T}{P} \frac{GM(r)\rho}{r^2},
  \label{eq:atmo_etransport_conv}
\end{align}
while radiative transport depends on the opacity $\kappa$ and the luminosity
$L$,
\begin{align}
  \frac{dT}{dr} 
   =  - \frac{3}{64\pi\sigma} \frac{\kappa L\rho}{r^2T^3}.
   \label{eq:atmo_etransport_rad}
\end{align}
The Stefan-Boltzmann constant is denoted by $\sigma$. 
The equation of state (EoS)
\begin{align}
  P = \frac{k_{\rm B}}{\mu} \rho T,
  \label{eq:EoS}
\end{align}
with $k_{\rm B}$ the Boltzmann constant and $\mu$ the mean molecular weight,
relates the pressure to the density and closes the system of equations.

In principle, one could solve for an energy equation, 
\begin{align}
  \frac{dL}{dr}=4 \pi r^2\rho \epsilon 
  \label{eq:atmo_energy}
\end{align}
where $\epsilon$ is the heat deposited at radius $r$. 
However, potential energy of accreted material is deposited deep in the
convective interior close to the core surface, so we take a
constant luminosity as function of planetary radius $L(r)=L$. 
When pebbles settle with terminal velocity in the
atmosphere, drag counterbalances gravity and locally deposits frictional heat
per unit mass
\begin{align}
  \delta E \approx \frac{GM_{\rm c}}{r^2} \delta r.
  \label{eq:drag_settle}
\end{align}
Therefore, per unit length, the deposited energy is much larger close to the
core surface than in the upper atmosphere, by a factor $10^6$ for the
atmospheres studied here. The luminosity profile takes the form
\begin{align}
  L(r) \approx \frac{GM_{\rm c}\dot M}{r_{\rm c}} - \frac{GM_{\rm c}\dot
  M}{r} = \left( 1-\frac{r_{\rm c}}{r} \right) L,
  \label{eq:lum_profile}
\end{align}
revealing that only near the core surface the luminosity deviates
from the constant value adopted here \citep{Rafikov_2006}.
Additionally, we ignore the heat from the contraction of the envelope, the
latent heat from evaporation and nuclear heating by the core through the decay
of $^{26}$Al, for the following reasons.
The luminosity generated from binding the gas envelope to the growing core can
be ignored when the core is subcritical \citep{Rafikov_2006}.
We find that latent heat of water sublimation can only be important for small
cores. 
For a certain accretion rate, we can assume that of the accreted material a
fraction $\beta$ of refractory grains settles to the core, while a remaining
fraction $1-\beta$ is water ice that sublimates.
The latent heat per unit mass required for the sublimation of the water
ice fraction is given by
\begin{align}
  Q_{\rm sub} = - 2.3 \times 10^{3} (1-\beta){\rm\,J\,g}^{-1}. 
  \label{eq:qsub}
\end{align}
The fraction that settles to the core gives
\begin{align}
  Q_{\rm grav} = \beta \frac{GM_{\rm c}}{R_{\rm c}}
  = 6.3\times 10^4 \beta \frac{M}{{\rm M}_{\rm E}} {\rm\,J\,g}^{-1}.  
  \label{eq:qgrav}
\end{align}
So the latent heat becomes important for $\beta=0.1$ and small cores ($M_{\rm
c} \lesssim 1$\,M$_{\rm E}$).
Finally, radioactive decay of short-lived radio isotopes in chondritic
material from the core releases a luminosity
$L = 1.5\times 10^{24} (M_{\rm
rock}/{\rm M}_{\rm E}) \exp\left( -t/\tau_{26{\rm Al}} \right)$\,erg/s. Here,
$\tau_{26Al}=1.01$\,Myr is the decay time of $^{26}$Al.
The importance of this heat source depends on the time of giant planet
formation after the formation of CAI, but remains about 4 orders of magnitude
smaller that the heat released by the accretion of pebbles.

In practice we integrate stepwise from the Hill sphere, where we assume nebular
conditions $(T_0,\rho_0)$, to the core surface in order to calculate the
envelope structure.
We iterate this procedure to take the self-gravity of
the envelope into account until we converge to a self-consistent solution.
Additionally, we take into account the sublimation of ice from settling
pebbles, by altering the mean molecular weight and the equation of state below
the ice line, under the assumption that convection causes an approximately
homogeneous mixture. There the molecular weight,
\begin{align}
  \mu_{\rm mix}^{-1} = \frac{1-X}{\mu_{\rm H/He}} + \frac{X}{\mu_{ {\rm
  H}_2{\rm O}}},
  \label{eq:mu_mix}
\end{align}
depends on the mass fraction of water vapour with respect to the Solar Nebula
H/He mixture $X$, with $\mu_{\rm H/He} = 2.34\,m_{\rm H}$ and $\mu_{ {\rm
H}_2{\rm O}} = 18\,m_{\rm H}$ ($m_{\rm H}$ is the mass of the H atom).
In the convective interior, the calculation of the temperature gradient relies
on the specific heat capacity of the
mixture $c_{\rm P,mix}= (1-X)c_{\rm P,H/He} + X c_{\rm P, {\rm H}_2{\rm O}}$,
with $c_{\rm P,H/He} = [\gamma_{\rm H/He}/(\gamma_{\rm H/He}-1)]k_B/\mu_{\rm
H/He}$ and  $c_{\rm P,{\rm H}_2{\rm O}} = [\gamma_{{\rm H}_2{\rm
O}}/(\gamma_{ {\rm H}_2{\rm O}}-1)] k_B/\mu_{ {\rm H}_2{\rm O}}$. 
We find that the critical core mass is very sensitive to the mean molecular
weight, but less so to the adiabatic index, which we we have taken here to be 
$\gamma_{\rm H/He} = 1.4$ (appropriate for a diatomic gas) and $\gamma_{{\rm
H}_2{\rm O}}=1.17$ (for water steam at high $T$ with all 12 degrees of freedom
released). More detailed calculations of the adiabatic index would require
solving for multiple chemical species in the envelope.
Such calculations show that in the limit of very polluted envelopes ($\sim
90$\,\%), from accreted material with a comet-like composition, changes in
$\gamma$ can lead to reduced critical core masses by at most a factor 2
\citep{Hori_2011}, which is an effect also seen in our simplified
model.

At wide orbital distances where the density is low, the photosphere is
located below the Hill sphere and the region is nearly isothermal with
$T^4 = T_0^4 + L/(16 \pi \sigma r^2)$\,\citep{Rafikov_2006}.
Examples of the envelope structure of planets located at various orbital
distances can be inspected in Fig.\,\ref{fig:rhoT_0510} and
Fig.\,\ref{fig:rhoT_2030}.
We further discuss the prescription of the opacity and the role of dust grains
in the next section.

\subsection{Determining the critical core mass}
\label{sec:critmasscalc}
We find the critical core mass numerically by stepwise increasing
the core mass.
When we not longer find a hydrostatic solution {for the envelope}, we identify
this core mass as the critical core mass, as is standardly
done \citep{Mizuno_1980}. 
The precise value of the critical core mass depends on
the assumed accretion rate, opacity and composition of the envelope. 

We have performed two classes of calculations of the critical core mass. 
In the first class, we have kept the accretion rate constant during the
iteration over the core masses, in order to decouple the stability of an
envelope from the assumed accretion rates by either planetesimals or pebbles
onto cores. 
The results are the black curves in Fig.\,\ref{fig:mvsmdotcrit}, where we
explicitly showed the dependency on the assumed composition of the envelope. 
In the second class, we have self-consistently taken into account the
dependency of the mass accretion rate of pebbles, and thus the luminosity, on
the mass of the planet. 
This is important in obtaining the results displayed in Fig.\,\ref{fig:ssfit}
and  Fig.\,\ref{fig:ssfit_1030}, where we calculate the planetary composition
as function of mass. 
The curves in these figures are obtained by considering two
regimes. In the first regime, the core has not reached isolation.
Consequently, for a given mass, and thus accretion rate, we find by iteration
the level of enrichment in heavy elements in the atmosphere required to collapse the envelope. 
The composition of the planet at this critical point is the one displayed by
the ratio of the heavy elements, in the core plus envelope, to the total planet
mass. 

In the second regime, the planet has grown beyond the isolation mass. In this
case, the mass of heavy elements is taken to be equal to that found in the
previous regime for a planet mass equal to the pebble isolation mass. 
The remainder of the mass of the planet is H/He from the gas accretion phase. 
Formally, at isolation the envelope does not have to be polluted to start gas
accretion, so one could also choose to only include the heavy elements from the
core, but this would constitute only a minor correction to the composition
compared to the total mass of the giant planets.

\subsection{Dependence on the (bulk) composition of the accreted material and
opacity}
\label{sec:composition}
We assume the bulk composition of the material accreted by the planet to
correspond to the bulk composition of cometary material \citep{Mumma_2011}.
The high fraction of water ice ($\approx$$50\%$) implies that the envelope gets
efficiently polluted by water vapour. 
Already at temperatures higher than $\approx$$100$\,K\,\citep{Supulver_2000}
ices sublimate, and this temperature  depends only weakly on the pressure.

The composition of the accreted pebbles also influences the opacity of the
envelope.
Icy grains set the optical depth at temperatures below $\sim$$100$\,K and below
the sublimation temperature of $\sim$$1000$\,K, opacity by silicate grains
dominates \citep{Pollack_1994}.
It is however not known how many grains are continuously deposited in the
planetary envelope after ice sublimation and through gas accretion from the
disc. Similarly it is poorly understood how fast grains settle, which depends
on their size, in turn set by the efficiency of grain growth and
fragmentation.

\begin{figure}
  \centering
  \includegraphics{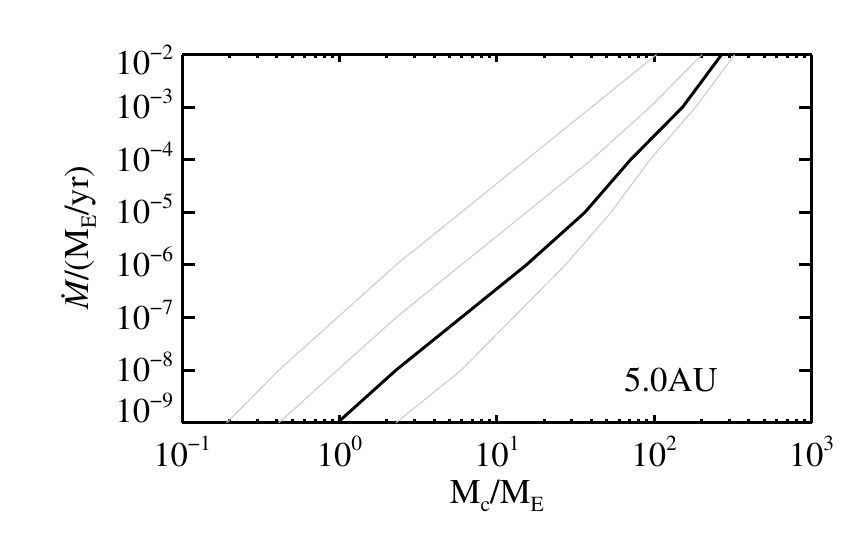}
  \caption{
  The dependency of the critical core mass on the dust opacity in H/He
  envelopes. 
  Grey curves correspond to factor 10 different dust opacities to the opacity
  used in this study (solid black line).
  The critical curves are here given at $5$\,AU, but have little dependency on
  orbital radius.
  }
  \label{fig:optest}
\end{figure}

Early core accretion studies \citep{Mizuno_1980} already pointed out the large
role the opacity of the envelope can play \citep{Ikoma_2000}. We have adopted a
modified version of the analytic expression for the Rosseland mean opacity by \citet{Bell_1994}.
Alternatively, one can approximate the opacity as a simple power law of
both temperature and pressure \citep{Rafikov_2006, Piso_2013}.
The black curve in Fig.\,\ref{fig:optest} corresponds to the
opacity used in this study, where we assume the dust opacity is reduced by a
factor 10 with respect to the disc \citep{Rafikov_2006}.
This choice reproduces the critical core masses (within a factor of 2) found by
\citet{Ikoma_2000}.

\section{Envelope contraction}
\label{sec:KH}

Finally, we briefly discuss the Kelvin-Helmholtz time-scale on which envelopes
contract
\begin{align}
  \tau_{\rm KH} \approx \frac{GM_{\rm c}M_{\rm env}}{R_{\rm eff}L}
  \label{eq:KH}
\end{align}
with $R_{\rm eff}$ some effective radius, typically taken to be the radius at
the convective to radiative border \citep{Ikoma_2000,Pollack_1996}.
After pebble isolation, the high luminosity from pebble accretion will almost
instantaneously contract the core to a state where only luminosity caused from
envelope contraction is important. 
Following numerical results by \citet{Ikoma_2000} contraction further occurs
on a timescale: 
\begin{align}
  \tau_{\rm KH}  = 3\times10^{5} 
  \left( \frac{M_{\rm c}}{10\,{\rm M}_{\rm E}} \right)^{-2.5} 
  \left( \frac{\kappa}{1\,{\rm cm}^2\,{\rm g}^{-1}} \right)^1
  {\rm yr,}
  \label{eq:ikoma_contract}
\end{align}
for envelope masses comparable to core masses $M_{\rm c}$.
The dependency on the dust opacity $\kappa$ means this is an upper limit in the
case of pebble accretion, as after isolation few grains will be further
deposited in the envelope.
Contraction could be further delayed by continued solid accretion
from planetesimals \citep{Pollack_1996}. However, this can be ignored in the scenario
described in the main paper, as firstly we do not assume all solid density in
planetesimals, and additionally we do not propose a significant enrichment
of the planetesimal column density (larger than a factor 5) as in
\citet{Pollack_1996}.
We emphasize that at the brink of collapse the Hill sphere of a gas giant is
comparable to the gas disc scale height, which makes the spherical approximation
invalid. 
Future studies will have to explore the effects of moving away from the
standardly assumed spherical symmetry.

\end{document}